\documentclass{aastex631}
\usepackage{hyperref}

\usepackage[utf8]{inputenc}

\begin{document}

\shorttitle{Small transient brightenings in moss region}
\shortauthors{Ram et al.}

\title{Transition Region Brightenings in a Moss Region and their Relation with Lower Atmospheric Dynamics}

\correspondingauthor{Tanmoy Samanta}
\email{tanmoy.samanta@iiap.res.in}

\author[0009-0007-2093-2795]{Bhinva Ram}
\affiliation{Indian Institute of Astrophysics, Koramangala, Bangalore 560034, India}
\affiliation{Max-Planck Institute for Solar System Research, 37077 Göttingen, Germany}

\author[0000-0002-9667-6392]{Tanmoy Samanta}
\affiliation{Indian Institute of Astrophysics, Koramangala, Bangalore 560034, India}

\author{Yajie Chen}

\affiliation{Max-Planck Institute for Solar System Research, 37077 Göttingen, Germany}

\author{Alphonse Sterling}
\affiliation{NASA Marshall Space Flight Center, Huntsville, AL 35812, USA}

\author{Jayant Joshi}
\affiliation{Indian Institute of Astrophysics, Koramangala, Bangalore 560034, India}

\author{Vasyl Yurchyshyn}
\affiliation{Big Bear Solar Observatory, New Jersey Institute of Technology, 40386 North Shore Lane, Big Bear City, CA 92314 9672, USA}

\author{Lakshmi Pradeep Chitta}
\affiliation{Max-Planck Institute for Solar System Research, 37077 Göttingen, Germany}

\author{Vaibhav Pant}
\affiliation{Aryabhatta Research Institute of Observational Sciences, Beluwakhan, 263001, Uttarakhand, India}

\begin{abstract}

Small-scale Brightenings (SBs) are commonly observed in the transition region that separates the solar chromosphere from the corona. These brightenings, omnipresent in active region patches known as “moss” regions, could potentially contribute to the heating of active region plasma. In this study, we investigate the properties of SB events in a moss region and their associated chromospheric dynamics, which could provide insights into the underlying generation mechanisms of the SBs. We analyzed the data sets obtained by coordinated observations using the Interface Region Imaging Spectrograph and the Goode Solar Telescope at Big Bear Solar Observatory. We studied 131 SB events in our region of interest and found that 100 showed spatial and temporal matches with the dynamics observed in the chromospheric H$\alpha$  images. Among these SBs, 98 of them were associated with spicules that are observed in H$\alpha$  images. Furthermore, detailed analysis revealed that one intense SB event corresponded to an Ellerman Bomb (EB), while another SB event consisted of several recurring brightenings caused by a stream of falling plasma. We observed that H$\alpha$  far wings often showed flashes of strong brightening caused by the falling plasma, creating an H$\alpha$  spectral profile similar to an EB. However, 31 of the 131 investigated SB events showed no noticeable spatial and temporal matches with any apparent features in H$\alpha$  images. Our analysis indicated that the predominant TR SB events in moss regions are associated with chromospheric phenomena primarily caused by spicules. Most of these spicules display properties akin to dynamic fibrils.
\end{abstract}

\section{Introduction}

The solar atmosphere is dominated by intense magnetic activities where strong heating events occur throughout the different layers of the atmosphere. Small-scale Brightenings (SBs), ranging in size from sub-arcseconds to a few arcseconds, are common features in the transition region (TR) images of the Sun \citep{Tian_2014_above_sunsopts,Skogsrud_2015,2015ApJ...803...44M} and could be potential sources of the heating of the upper solar atmosphere. Among these dynamic SBs, some appear roundish in shape, while others are thin elongated jet-like features. Although SBs with a variety of sizes and intensities are observed in different regions of the Sun, the mechanisms responsible for generating these SBs remain unclear.

In quite-Sun and coronal hole regions, chromospheric spicules \citep{1968_Beckers,1982_Athay,2000_Sterling,2011_spicule_corona_Pontieu,2012_Tsiropoula} are suggested to be a dominant source of fast-moving jet-like SBs in the TR network, often called network jets \citep{2014_tianb,2015ApJ...803...44M,2016_Narang}.
Rapid Blue-shifted Excursions (RBEs) are chromospheric jets observed as absorption features in far-blue wings of H$\alpha$ and Ca II IR lines due to their high line-of-sight velocities \citep{Langangen_2008c, 2009_Rouppe}. They are believed to be on-disk counterparts of the high-speed chromospheric jets, so-called type-II spicules \citep{2007b_de_Pontieu} and are often related to TR network jets \citep{Rouppe_van_der_Voort_2015}.
These spicules are suggested to be generated by the reconnection process and often cause heating of more than transition region temperatures \citep{2014_Pereira,Rouppe_van_der_Voort_2015,2019_Samanta}
which mostly appears as SB in the TR.
SB events are also abundantly seen in internetwork regions of quiet-Sun and coronal holes.
\cite{2015ApJ...803...44M} suggested that acoustic shocks could generate these SB events in the internetwork regions. They proposed that upward-moving acoustic waves in a non-magnetic environment get converted into shocks due to a steep decrease in the density, resulting in heating.

The TR SB events are pervasive in active regions, including above sunspots.
Several studies investigated SB events above sunspots \citep{Tian_2014_above_sunsopts, 2015ApJ...811L..33V,Deng_2016,2017ApJ...835L..19S}. Most of these TR brightening events within sunspots
are short-lived and small in size.
Coordinated observations of chromospheric and transition region above sunspots \citep{Tian_2014_above_sunsopts,2016ApJ...816...92T,2017ApJ...835L..19S} suggest that some of the TR SBs in penumbrae are heating signatures of chromospheric penumbral microjets \citep{2007Sci...318.1594K}. It is suggested that the TR SBs could be generated by magnetic reconnection \citep{2016ApJ...816...92T} or due to falling plasma \citep{2017ApJ...835L..19S}.

Ellerman bombs (EBs), UV Bursts, and Explosive Events often produce SBs around active regions; however, they generally create ``intense'' SB events in the TR \citep{1989_Dere_EEs,1998_Ding_EB_mustach,2014_H_Peter_EE,2017_Chitta_UVB,2018_Young_UVB,2019_Chen_UVB}. These events are produced by magnetic reconnection at different heights in the atmosphere.
EBs are small-scale brightening events that appear in the wings of H$\alpha$, and Ca II 8542~{\AA} \citep{1917ApJ....46..298E,2002ApJ...575..506G, Rutten_2013}. EBs generally show a significant intensity enhancement in the H$\alpha$ wings without any noticeable intensity enhancement in the core which creates a mustache-like H$\alpha$ spectral profile. They are believed to be caused by magnetic reconnection between oppositely oriented field lines of U-shaped loops \citep{2008_Watanabe,2012_Pariat}.
Recently, it was suggested the UV Bursts \citep{2014_H_Peter_EE} can also be generated by a similar reconnection process as EBs; however, the reconnection height may be higher in the solar atmosphere \citep{2019ApJ...875L..30C}.
Explosive events result from busty reconnection at transition region heights, producing strong flows within the transition region, which generally affects the transition region spectral line profiles as well as produces localized intensity enhancement \citep{chen_2019_TR_EEs}.

 Interface Region Imaging Spectrograph \citep[IRIS;][]{2014_IRIS_Pontieu} observations of active regions revealed short-lived bright grain-like features of size $\sim0$\farcs{5} and lifetime $\sim$3 minutes at transition region temperatures \citep{2016ApJ...817..124S}. The study suggested that TR SBs (bright grains in their paper) in plage regions result from acoustic shocks that drive dynamic fibrils. Dynamic fibrils are fine, thread-like structures mostly observed near the H$\alpha$ line core in active region plages \citep{2007ApJ...655..624D}. They usually follow parabolic paths with a symmetrical ascending and descending phase, typically having speeds of around 15-20 km $s^{-1}$.  These structures are believed to form due to evanescent photospheric $p$-modes leaking into the atmosphere through inclined magnetic fields, leading to shock formation, and are commonly known as type-I spicules.
 Recently, coronal signatures of these chromospheric dynamic fibrils have also been identified by using high-resolution observations from the Extreme Ultraviolet Imager \citep[EUI;][]{2020A&A...642A...8R} aboard the Solar Orbiter \citep{2023A&A...670L...3M, 2023A&A...678L...5M}.
 On the other hand, recent active region observations by \citet{2022MNRAS.515.2672V} showed that many RBEs are related to TR SB events in the plage regions.
 Note the presence of RBEs and their association with transition region emission are generally investigated in quite-sun and coronal hole regions \citep{2014_Pereira,2015_Rouppe_van_der_Voort} and rarely reported in active regions.
 These investigations clearly show that the exact generation mechanism of SB events is not settled.
 Therefore, it is crucial to delve into more high-resolution data and perform statistical analyses to gain a better understanding of the relationship between SBs and chromospheric dynamics.

We note that several detailed studies of ``intense'' brightening events have been conducted in recent years in active regions \citep{2018_Young_UVB}. However, using high-resolution data, only a few studies investigated the most abundant and faint SBs in detail within active region ``moss''. Moss regions are finely textured and low-lying EUV emission patches above active region plage \citep{2000_Martens,2010_Tripathi}. This low-lying emission was suggested to be coming from the locations of the footpoints of coronal loops in the dense cores of active regions \citep{1999_Berger,2003_Antiochos,2010_Tripathi} where many SBs are observed in the transition region.
We can now resolve incredibly small brightening features in the lower solar atmosphere and study their contributions to the heating of transition region and corona with the progress and developments in observational solar physics such as the 1.6 m Goode Solar Telescope \citep[GST;][]{2010_Cao} at Big Bear Solar Observatory (BBSO), IRIS, and Atmospheric Imaging Assembly \citep[AIA;][]{2012_AIA_Lemen} on-board Solar Dynamics Observatory. In this article, we explored the properties of TR SB events in a moss region and their chromospheric connection in detail using high-resolution IRIS and GST data. \\

\begin{figure*}

 \centering
\includegraphics[width=0.58\textwidth,angle=90]{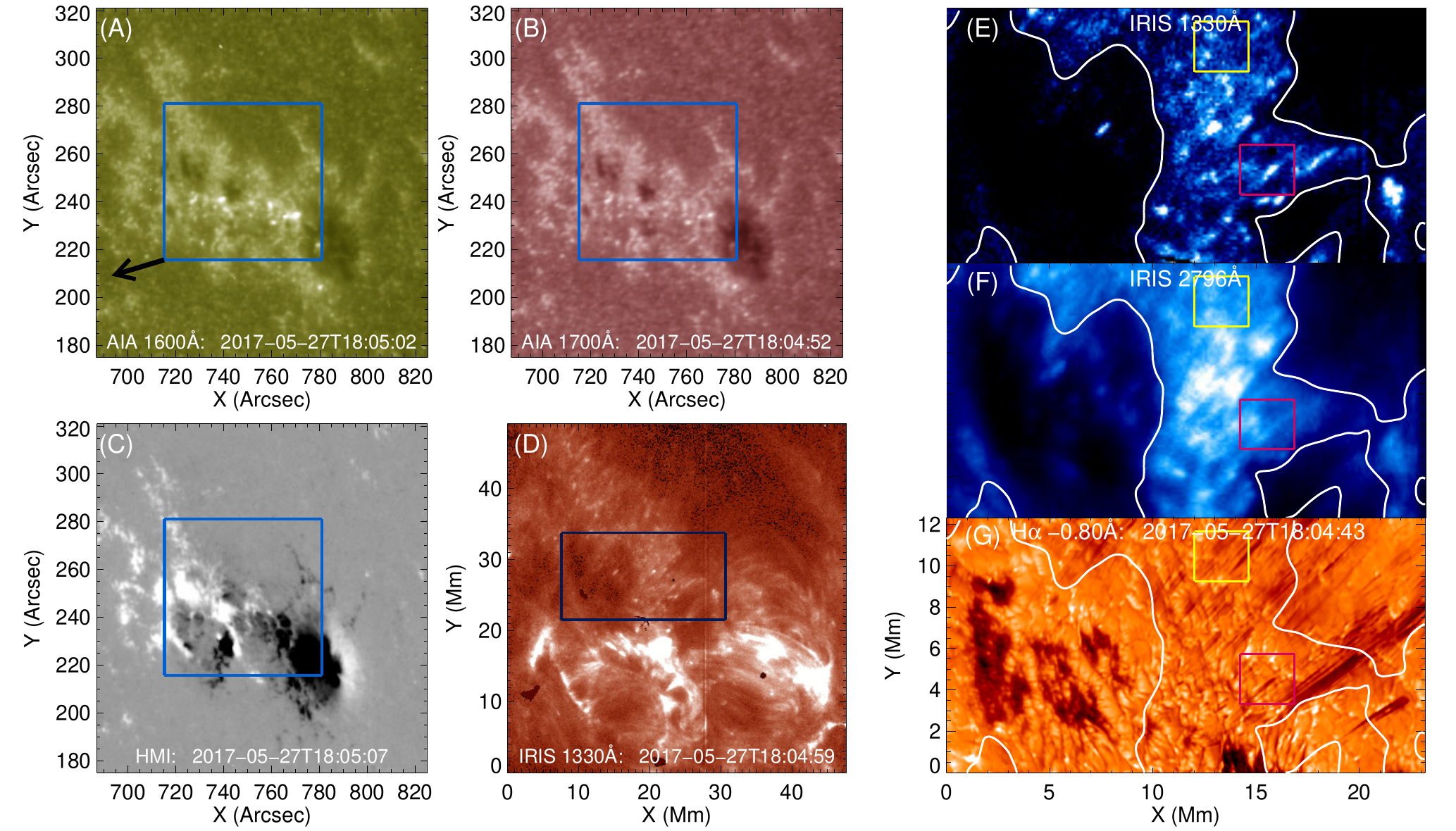}
 \caption{(A)-(B): Show the plage (moss) region in AIA 1600~{\AA} and 1700~{\AA} filters respectively at 18:05:02 UT on 2017 May 27. (C): Shows the HMI line-of-sight magnetic field map of the same region. The black arrow in panel (A) points toward disc center. The pale blue rectangle in panels (A)-(C) shows the Field-Of-View (FOV) of SJI 1330~{\AA} in panel (D). Many TR SB events are clearly observed in the SJI 1330~{\AA} (panel D). The dark blue rectangle marks our region of interest and is shown in panels (E)-(G) in different filters as marked.
The white contours, derived from a smoothed SJI 1330~{\AA}, represent the boundary of the moss region in SJI 1330~{\AA}. The red square represents the FOV shown in Figure \ref{fig:cor}, where we present examples of SBs in SJI 1330~{\AA} that clearly match with spicules in the wings of H$\alpha$. The yellow square corresponds to the field of view of Figure \ref{fig:noncor}, which shows an example of brightening in 1330~{\AA} images that do not match with any apparent dynamic features in H$\alpha$ images. Note that H$\alpha$ images in all the figures are displayed in the logarithmic scale. An animation of this figure is available \href{https://www.dropbox.com/scl/fi/n005yg3h7e8yh2ywk4qxs/M1.mp4?rlkey=or83xb8dg4lhag0cochmymki4&dl=0}{(M1)}, presenting the entire FOV as seen by IRIS 1330~{\AA} filter throughout the complete observational period. The animation includes the Region-Of-Interest (ROI) highlighted in panels (E), (F), and (G), providing an overview of the ROI.}
 \label{fig:ref}
\end{figure*}

\section{Observations}

We analyzed the data recorded by coordinated observations between the 1.6m GST at BBSO and the IRIS on May 27, 2017. IRIS performed a 16-step raster of NOAA AR 12659 at a step size of $2\arcsec$, covering $120\arcsec$ along the slit. It provided 2832~{\AA}, 2796~{\AA} and 1330~{\AA} Slit-Jaw Images (SJIs) with cadences of 83~s, 21~s and 21~s, respectively with a plate scale of $\sim$0\farcs{165} pix$^{-1}$. The emission in the IRIS 2796~{\AA} filter is dominated by the emission from the upper chromosphere ($\sim10^4$K), whereas the IRIS 1330~{\AA} filter is sensitive to lower transition region temperatures ($\sim10^{4.4}$K).
The Visible Imaging Spectrometer (VIS) of GST recorded narrow-band images of the H$\alpha$ line core and H$\alpha$ wings at ±1, ±0.8, ±0.6, ±0.4, and ±0.2~{\AA}, alternately, with a spatial pixel size of $\sim$0\farcs{029} and a cadence of 53~s approximately. The H$\alpha$ line core generally forms at the chromospheric heights, whereas the wings form at lower atmospheric heights \citep{2006A&A...449.1209L,2012ApJ...749..136L}. We used 2832~{\AA} SJIs to co-align the IRIS SJIs with the GST data set by comparing common photospheric features with H$\alpha$ -1~{\AA} wing images for each frame.
Note that the mentioned data sets were previously utilized by \cite{2019ApJ...875L..30C}, and the detailed data reduction techniques have already been outlined there. We analyzed the recorded data during 17:50-19:24 UT, and our ROI ranged from $720\arcsec$ to $745\arcsec$ solar X and from $245\arcsec$ to $263\arcsec$ solar Y.
It is worth mentioning that the active region we are studying exhibits intense events like large jets and flares. Therefore, we selected the ROI based on the availability of GST data and to avoid strong magnetic activities (e.g., flare, large brightenings, larger jets, etc.) which allowed us to concentrate on studying the SB events at active region moss. Furthermore, note that AIA and Helioseismic and Magnetic Imager \citep[HMI;][]{2012_Schou_hmi} data onboard SDO satellite are used for reference purposes and clearly identifying the active region plage (see Figure~\ref{fig:ref}).

\section{Data Analysis and Results}

 This study focuses on examining TR SB events in a moss region and their chromospheric connections (Figure~\ref{fig:ref}). The moss regions around the sunspots are clearly visible in the AIA 1600~{\AA}, 1700~{\AA} passbands, and SJI 1330~{\AA}. Figure~\ref{fig:ref} and the associated animation show many dynamic TR SB events visible in the IRIS SJI 1330~{\AA} within the moss regions. We have mostly focused on the small part of the moss region marked by the white contours where most of the SB events appear.
Initially, we visually investigated these TR SB events and noted that spicules (dark collimated jets observed in H$\alpha$ wings) are a common occurrence in this region. We selected those SBs that are observed in three or more consecutive 1330~{\AA} images. Afterward, we searched for any corresponding chromospheric counterparts of these SBs. We found that a few SBs were observed at locations where numerous chromospheric features were present simultaneously, making it challenging to identify the relevant ones for the respective SBs accurately. Consequently, we excluded those SBs from our study. Based on these criteria, 131 SB events were identified in our ROI and selected for further analysis.
Finally, we focused on exploring the chromospheric counterparts of these SBs in detail.

\begin{figure*}
 \centering
 \includegraphics[width=0.57\linewidth,angle=90]{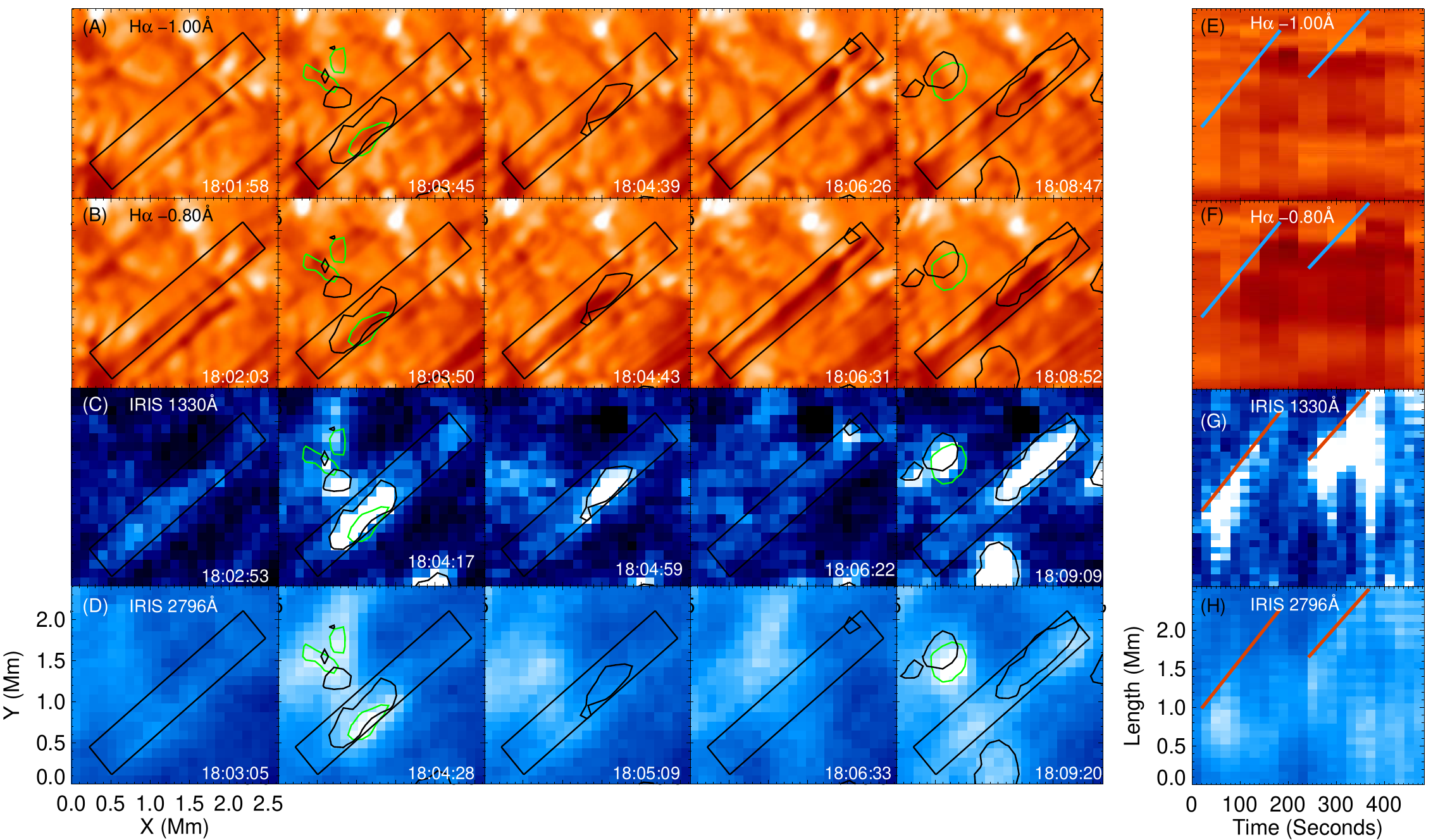}
 \caption{Time evolution of TR SB events which clearly match with the evolution of two spicules. Panels (A)-(D) show the FOV of the red square in panels (E)-(G) of Figure \ref{fig:ref}. (A) and (B): Time evolution of two spicules in H$\alpha$ -1.00~{\AA} and -0.80~{\AA}, respectively. (C) and (D): Time evolution of two SB events in SJI 1330~{\AA} and 2796~{\AA}, respectively. The black and green contours correspond to the SBs seen in SJI 1330~{\AA} and 2796~{\AA} images, respectively. The black rectangle corresponds to a 0\farcs{8} wide artificial slit taken to track the spicules and their counterparts in IRIS SJIs. The SBs (contoured) in SJI 1330~{\AA} can clearly be seen following the ends of the spicules in H$\alpha$ as they move upwards. SBs in SJI 2796~{\AA} only appear for a single frame, but faint brightening can be seen mapping the body of the spicules throughout its lifetime. (E)-(H): Space-time plots for the artificial slit shown as a black rectangle in panels (A)-(D), respectively. Inclined lines mark the motion of two SB events corresponding to two spicules. }
 \label{fig:cor}
\end{figure*}

\begin{figure*}
 \centering
 \includegraphics[width=\linewidth]{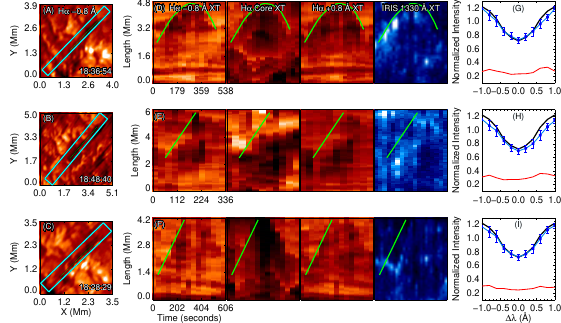}
 \caption{Examples of three spicules with different spectral profiles that have a correspondence with TR SB events. Panels (A)-(C) show the spicules as seen in H$\alpha$ -0.80~{\AA} wing. Panels (D)-(F) show their space-time diagrams for the cyan-colored artificial slits in panels (A)-(C). From left to right in each panel, we display the H$\alpha$ -0.80~{\AA}, line core, +0.80~{\AA} wing, and IRIS 1330~{\AA}, respectively. The spicule shown in panel (A) follows a parabolic trajectory which is marked by a green curve. The green line in panels (E) and (F) also marks the trajectory followed by the spicules as seen in H$\alpha$ -0.80~{\AA} wing. (G)-(I): The black curve shows the average H$\alpha$ spectral profile. The blue curve corresponds to the H$\alpha$ spectral profiles of the respective spicules calculated by averaging the intensity of the 3-pixel wide slit along the jet shown as a green line in panels (A)-(C). The error bars represent 1{$\sigma$} uncertainties in the measurements. The red curve shows the difference between the average spectral profile and the spectral profile of the spicule.}
 \label{fig:jet_sp_xt}
\end{figure*}

\begin{figure*}
 \centering
 \includegraphics[width=0.57\linewidth,angle=90]{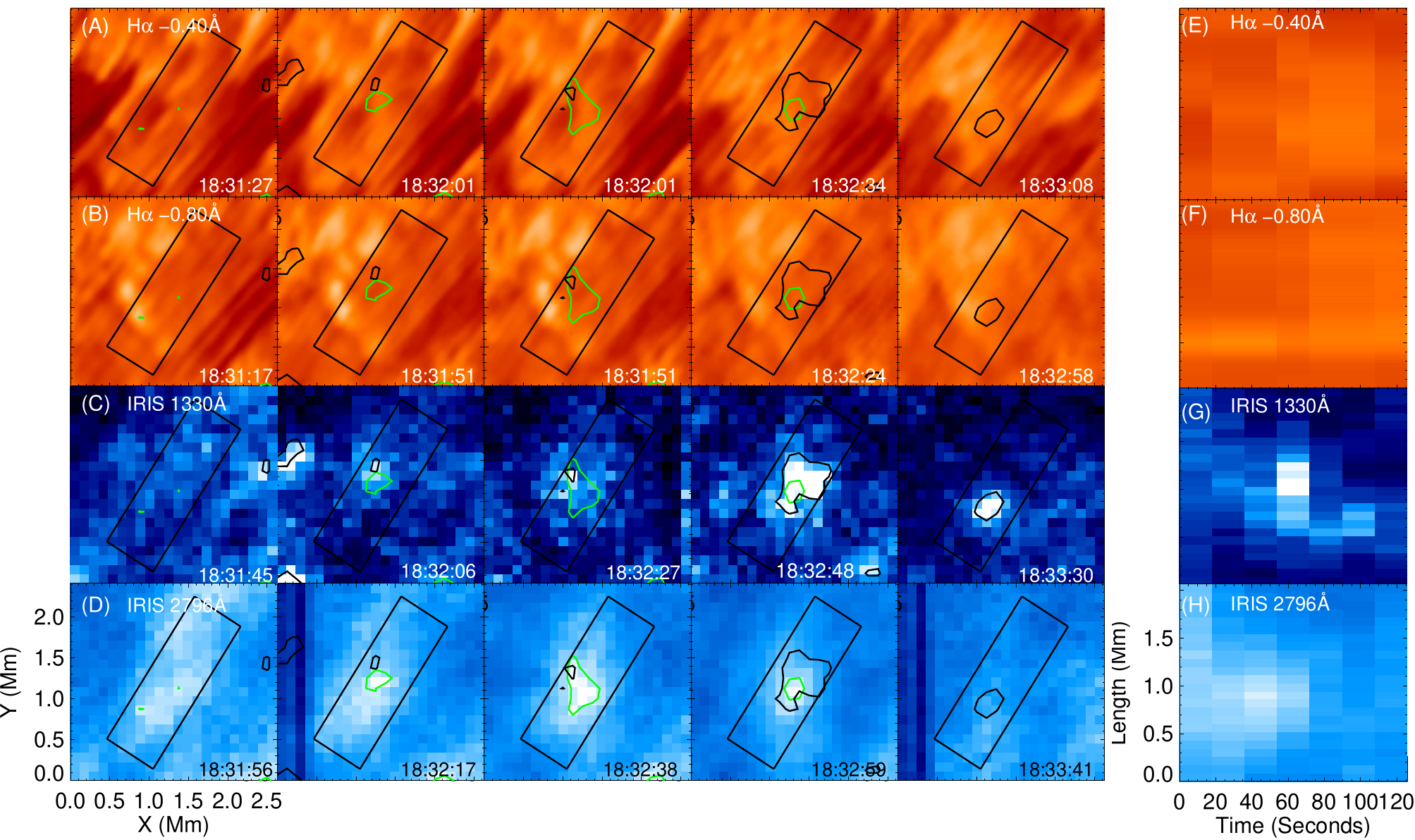}
 \caption{A typical example of an SB event that does not correspond with any apparent feature in H$\alpha$. The panels are the same as in Figure~\ref{fig:cor}. The SBs in SJI 1330~{\AA} and 2796~{\AA} (panels in (A)-(B)) do not show any counterparts. Space-time plots (E-H) also clearly show no signature of the IRIS SBs in H$\alpha$ wings.}
 \label{fig:noncor}
\end{figure*}

We first started investigating whether spicules could be a counterpart of some of the SBs. We define a good spatiotemporal match between an SB and a spicule when a spicule appears within a radius of $\sim$0\farcs{5} of the structure of an SB during its lifetime. This spatial offset between SBs and spicules can be caused by the projection effect since the ROI is located close to the limb.
Figure~\ref{fig:cor} shows an example of a clear spatiotemporal match between SBs and spicules. The location of this observation is marked in Figure~\ref{fig:ref}~(E)-(G) by a red-colored rectangle. The time evolution of the GST H$\alpha$ wings images and IRIS SJIs (panels A-D) shows the two SB events that occur at the top of spicules. We further produced space-time plots (panels E-H) to demonstrate this connection clearly. Note that the space-time plots of IRIS SJIs in Figure~\ref{fig:cor} (also in Figs.~\ref{fig:jet_sp_xt}, and \ref{fig:noncor}) are derived by taking the average intensity of all the pixels along the width of the slit. Additionally, the space-time plots of H$\alpha$ images are derived by taking the minimum intensity of all the pixels along the width of the slit. The spatiotemporal match between spicules and SB events in SJI 1330~{\AA} images is clearly visible in these space-time plots. Detailed analysis of all the events reveals that 98 SB events in SJI 1330~{\AA} images are related to spicules. These SBs show good spatiotemporal matches with spicules and typically follow their tips. We further notice that many of these SBs also appear in 2796~{\AA} images however, most of them map the middle part of these spicules (see bottom panels of Fig.~\ref{fig:cor}).\\

We conducted a more detailed investigation to determine whether these TR SB-related spicules can be classified as either dynamic fibrils or RBEs. Figure \ref{fig:jet_sp_xt} provides insights into the distinct characteristics of the spicules that are linked to SB events. Three rows in the figure depict three prevalent types of observational features that we identified through our analysis.  In order to comprehend their dynamics better, we present space-time diagrams in panels (D)-(F) and showcase their corresponding H$\alpha$ spectral profiles in panels (G)-(I).

The top panels illustrate an example of a spicule that follows a parabolic trajectory in the space-time plots, resembling dynamic fibrils \citep[e.g.,][]{2007ApJ...655..624D}. The H$\alpha$ spectral profile exhibits a slight broadening compared to the average profile. Notably, only a small fraction of these spicules (9 out of 98) display a clear parabolic trajectory in the H$\alpha$ space-time diagrams. The middle panel presents another example where the collimated jets fade away after reaching their apex, without a distinct indication of falling back. The spectral profile shows a slight broadening of the line without noticeable asymmetry. Although the abrupt vanishing of the jets may suggest characteristics of an RBE, the absence of blue/red wing enhancement in the H$\alpha$ line profile (as seen in the red curve in panel H) prevents us from categorizing these jets as RBEs. Conversely, the spicule in the bottom row exhibits an ambiguous trajectory in the space-time diagrams, along with a spectral profile similar to the previous two types with slight broadening.
These predominant spicules either fade upon reaching their peak heights or display trajectories that are challenging to classify. We perform detailed calculations on the kinematics of these spicules, including the deceleration of seven spicules with parabolic trajectories and the speeds of 15 spicules that fade after reaching their peak heights. These calculations are performed by tracing their trajectories in their space-time plots. The deceleration values presented here represent the average apparent deceleration shown by these spicules during the rising phase in H$\alpha$ core space-time plots, while the apparent speeds are the average speeds from the point they appear to when they fade away in H$\alpha$  -0.80~{\AA} wing space-time plots. The deceleration values for these spicules range from 40$~m~s^{-2}$ to 211$~m~s^{-2}$, while the apparent rise velocities vary between 5$~km~s^{-1}$ and 25$~km~s^{-1}$. These values are on the lower end of the range observed by \cite{2007b_de_Pontieu}.
In summary, this analysis indicates that these spicules generally demonstrate broadened profiles, decelerations, and speeds akin to those caused by dynamic fibrils.

Detailed analysis also revealed that several SB events do not have spatiotemporal correspondence with any dynamical
features in H$\alpha$ images.  Note that, consistent with our previous analysis, we searched for any noticeable spatiotemporal dynamics in H$\alpha$ images corresponding to an SB that may occur within approximately of $0\farcs{5}$ of the structure the SB's  radius during its lifetime. Figure \ref{fig:noncor}
highlights an example where the SB event does not show an apparent spatiotemporal match with a spicule or any other feature in H$\alpha$ wings. Note that the location of this event is also marked in Figure \ref{fig:ref} by a yellow-colored rectangle.
Figure \ref{fig:noncor} presents the time evolution of such an SB event in SJI 1330~{\AA} and 2796~{\AA} images (see panels C and D). Panels (A) and (B) clearly reveal that there is no apparent signature of the corresponding chromospheric features inside the black rectangle in the wings of H$\alpha$ (panels A-B). Panels (E)-(F) show the space-time plots of H$\alpha$ -0.40~{\AA} and -0.80~{\AA} wings, respectively, for the slit marked by the black rectangles in panels (A)-(D). Whereas, panels (G)-(H) are the space-time plots of SJI 1330~{\AA} and SJI 2796~{\AA} for the same slit.
The space-time maps also make it evident that there is no apparent sign of any interrelated features. We observed 31 such events (out of 131 investigated events) where SBs do not show any apparent spatial and temporal match with any visible features in H$\alpha$ wings, which suggests that either the corresponding features have a width below the resolution limit of our observational data or these SB events are entirely a TR phenomenon.

\begin{figure*}

 \includegraphics[width=0.53\linewidth,angle=90]{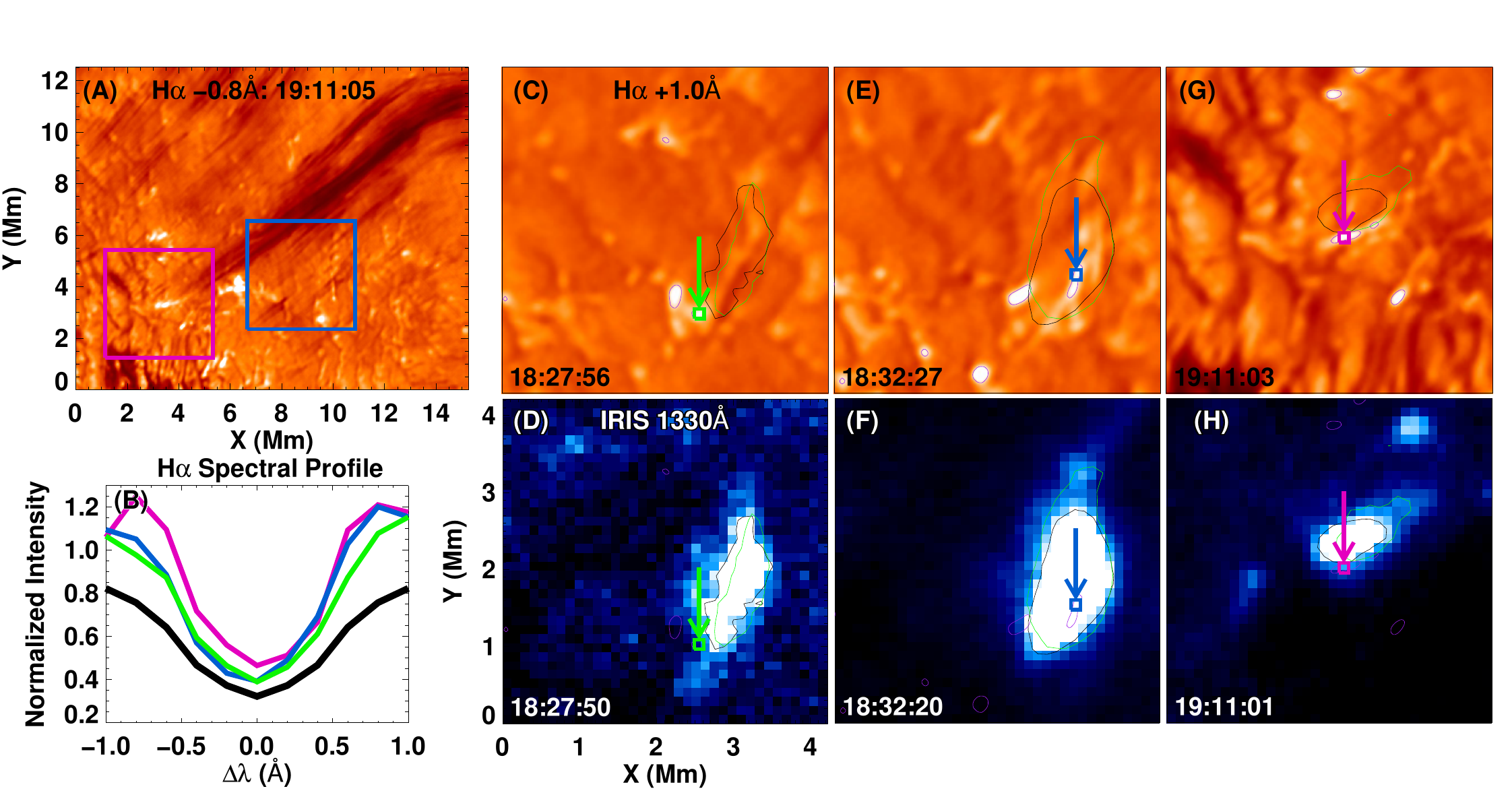}
 \caption{Relation of SBs with EB and falling plasma. (A): Large FOV covering the falling plasma and an EB in H$\alpha$ -0.8~{\AA} at 19:11:05 UT on May 27. The blue square corresponds to the FOV of panels (C) and (E), and the pink square outlines the FOV of panel (G). (B): H$\alpha$ spectral profile: the thick black line shows the average spectral profile, whereas green, blue, and pink lines correspond to the spectral profiles of the events shown in panels (C), (E), and (G), respectively. (C) and (E): Shows the EB in H$\alpha$ +1.00~{\AA}. (D) and (F): Shows the same FOV as panels (C) and (E), covering profound brightening near the EB in SJI 1330~{\AA}. Purple contours indicate 3$\sigma$ intensity enhancement above the mean in H$\alpha$ +1.00~{\AA} wing images. Black and green contours map the brightening in SJI 1330~{\AA} and 2796~{\AA} images respectively. (G): Showcases the brightening in H$\alpha$ +1.00~{\AA} images related to falling plasma. (H): SJI 1330~{\AA}, which shows SBs near the brightening in H$\alpha$ +1.00~{\AA}. green, blue, and pink arrows point to the location of the brightenings in H$\alpha$ +1.00~{\AA}. The green and blue arrows point to the brightening of an EB, whereas the pink arrow points to the falling-plasma-related-brightening. An animation for this figure is available (\href{https://www.dropbox.com/s/si37myifcey0ct3/m5.mp4?dl=0}{M5}) that includes panels (A) and (C)-(H), where the downflows and brightenings are clearly vissible .
 }
 \label{fig:fp_eb}
\end{figure*}

Our investigation reveals long-lasting intense recurring brightenings in SJI 1330~{\AA} and 2796~{\AA} in the neighborhood of the brightenings in H$\alpha$ far wings (panels C-D of Figure~\ref{fig:fp_eb}). For further analysis, we calculate H$\alpha$ spectral profiles of these bright features by taking the intensity of the brightest pixel inside the small squares (boxes of 7$\times$7 pixels) at which the arrows are pointing in panels (C), (E), and (G). The thick black line in panel (B) is the reference profile derived by taking the average intensity of a large region
for the entire observational period. The  H$\alpha$ spectral profiles (drawn in green and blue in panel B) corresponding to the SB event in panels (C) and (E) look like those of an EB. These profiles show a large intensity enhancement at the wings and remain faint in the core throughout the observational period. Additionally, these brightenings in the H$\alpha$ +1.00~{\AA} wing images exhibit an intensity enhancement exceeding 3$\sigma$ above the mean.
Using the same data, \cite{2019ApJ...875L..30C} also reported many similar events in this active region. They suggested that the brightenings in SJI 1330~{\AA} near EBs can occur if the reconnection current sheet extends well above the temperature minimum region. When the reconnection near the temperature minimum region generates EBs, the same reconnection current sheet at chromospheric heights can produce UV bursts. The spatial offset between SBs and EBs can be a consequence of this height difference and the projection effect since the region of interest is close to the limb. Therefore, the temporal match between these bright features becomes crucial to identify their corresponding features.
The EB events, often occurring in fragmented bursts, persisted throughout the entire duration, reappearing intermittently in the same vicinity. Sometimes, the SBs appear right above the EB. We measured the distance between the centers of the EBs and the SBs, finding offsets primarily in the radially outward direction, with SBs appearing above the EBs.  The example in panels (C-D) shows an offset of around 900 km between the EB and SB, while the EB and SB in panels (E-F) exhibit almost no offset. The distances during different occurrences varied from 0 to 1000 km, with a median of around 600 km. This range is consistent with the findings of \cite{2019ApJ...875L..30C}, who reported similar offsets in their study of EB and UV bursts.

 Our analysis also reveals recurring transient brightenings in SJI 1330~{\AA}, far wings of H$\alpha$, and sometimes in SJI 2796~{\AA} near the footpoints of a stream of falling plasma (see the animation associated with Fig.~\ref{fig:fp_eb}). These prominent H$\alpha$ wing brightenings appear to be similar to that reported by \citet{2016NatSR...624319J}, where the falling plasma along the flare loops causes strong brightenings at their footpoint. We investigated the H$\alpha$ spectral profiles of these events in detail. An example of the falling-plasma-related-brightening event is presented in panels (G) and (H) of Figure~\ref{fig:fp_eb}.
 The H$\alpha$ spectral profile at the region of this brightening is shown in pink in panel (B). We note that the H$\alpha$ spectral profile looks similar to an EB. However, sometimes, the line core also shows intensity enhancement along with the wings.
 The SBs in IRIS pass-bands appear co-temporally near the falling plasma regions. These SBs tend to move erratically along with their corresponding H$\alpha$ brightenings. SBs corresponding to these falling-plasma-related-brightening events have different shapes and sizes.
 The evolution of these events is extremely difficult to track due to the temporal resolution of the GST data set and the small sample size. Some of these events
 only appear for a few frames ($\sim$1-4) in H$\alpha$ images while others
 last throughout the entire observational period. Furthermore, some of the falling-plasma-related-brightening events merge with each other, whilst some of them split into two or more, making it difficult to track these individual events. Additionally, we estimated the speeds of the falling plasma by producing space-time plots along the falling streams and found that the falling plasma speeds range from 15 to 20 km s$^{-1}$.

\begin{figure*}
 \centering
 \includegraphics[width=0.52\linewidth,angle=90]{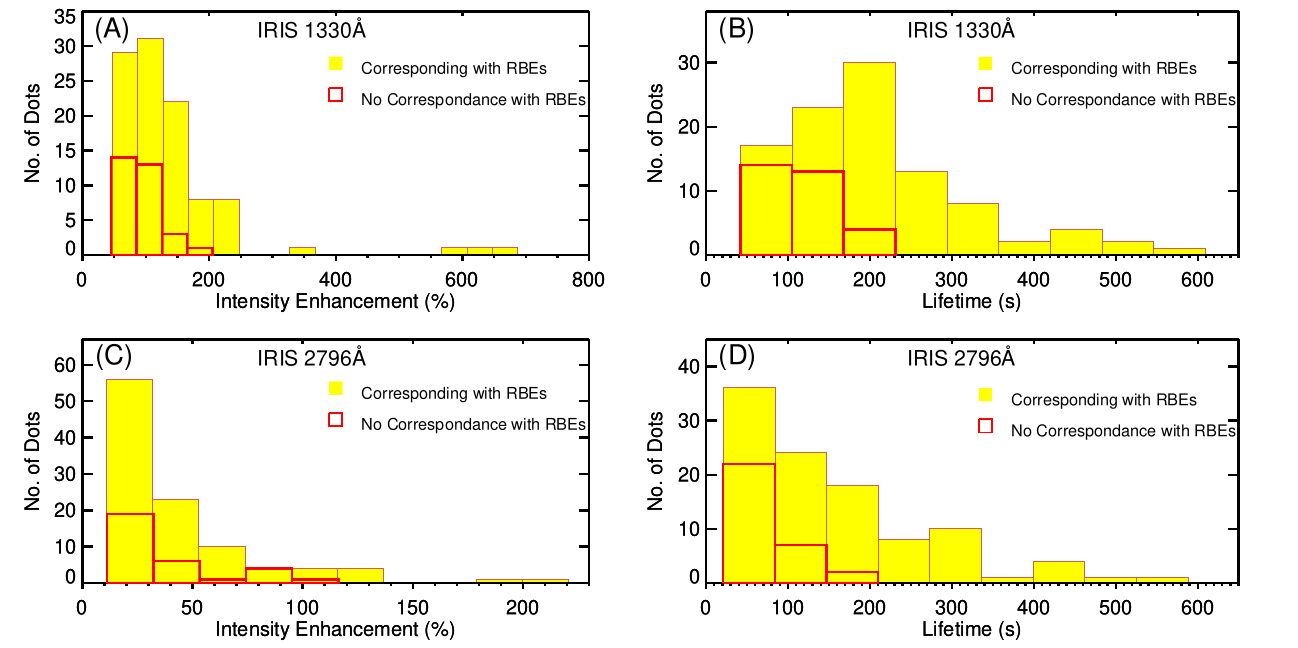}
 \caption{(A)-(B): Distribution of intensity enhancements and lifetimes of SB events in SJI 1330~{\AA}, respectively. (C)-(D): Distribution of intensity enhancements and lifetimes of SB events in SJI 2796~{\AA}, respectively. The solid yellow bins represent those SB events that match with the spicules while hollow red bins correspond to those SB events that do not match with any features in the H$\alpha$ images.
 }
 \label{fig:histo}
\end{figure*}

Our analysis showed that the majority of the SBs accompany spicules, however, there are many SBs where no associated dynamic changes are observed in the chromosphere. We compared the properties of these two different types of SB events, primarily focusing on their intensity enhancements relative to background intensity and their lifetime. To estimate these parameters, we utilized the slits in panels C-D in Figures~\ref{fig:cor} and \ref{fig:noncor}. By examining space-time diagrams of SB events (panels G-H in Figs.~\ref{fig:cor} and \ref{fig:noncor}), we determined their lifetime by setting an intensity threshold for each IRIS passband. The pixels exceeding this threshold in the frame were isolated, and their maximum intensities were recorded. Additionally, we calculated the average intensity below this threshold. The pixels exceeding this threshold in the frame were isolated, and their maximum intensities were recorded. Additionally, we calculated the minimum average intensity of the isolated pixels throughout the lifetime of an SB event, which served as the background intensity. The ratio of the maximum intensity to the average background intensity yielded the percentage of intensity enhancement attributed to each SB event.

Figure \ref{fig:histo} illustrates the distribution of intensity enhancements and lifetimes of SB events for both the 1330~{\AA} and 2796~{\AA} passbands.
The histograms clearly demonstrate that SB events exhibiting a co-spatial and co-temporal correspondence with spicules (depicted in yellow) are generally brighter and have longer lifetimes compared to those lacking spatial-temporal dynamics in the chromosphere (depicted in red). However, it is important to highlight that the characteristics of non-spicule-related SB events fall at the lower end of the spectrum compared to SB events caused by spicules.
It is plausible that these SB events represent a scaled-down version of spicule-related SB events, with spicules themselves potentially going undetected in H$\alpha$ images due to current observational limitations. Additionally, it is worth noting that SBs associated with Ellerman bombs (EB) and falling-plasma-related events exhibit significantly higher brightness and longer durations than the commonly observed SBs in this moss region. Specifically, the maximum intensity enhancement attributed to SBs due to falling plasma reached 522\%\ while the intensity enhancement for the SB event associated with EB was as high as 1700\%\ in SJI 1330~{\AA} images.

\section{Discussion}

\begin{figure}
 \centering
 \includegraphics[width=0.63\linewidth,angle=90]{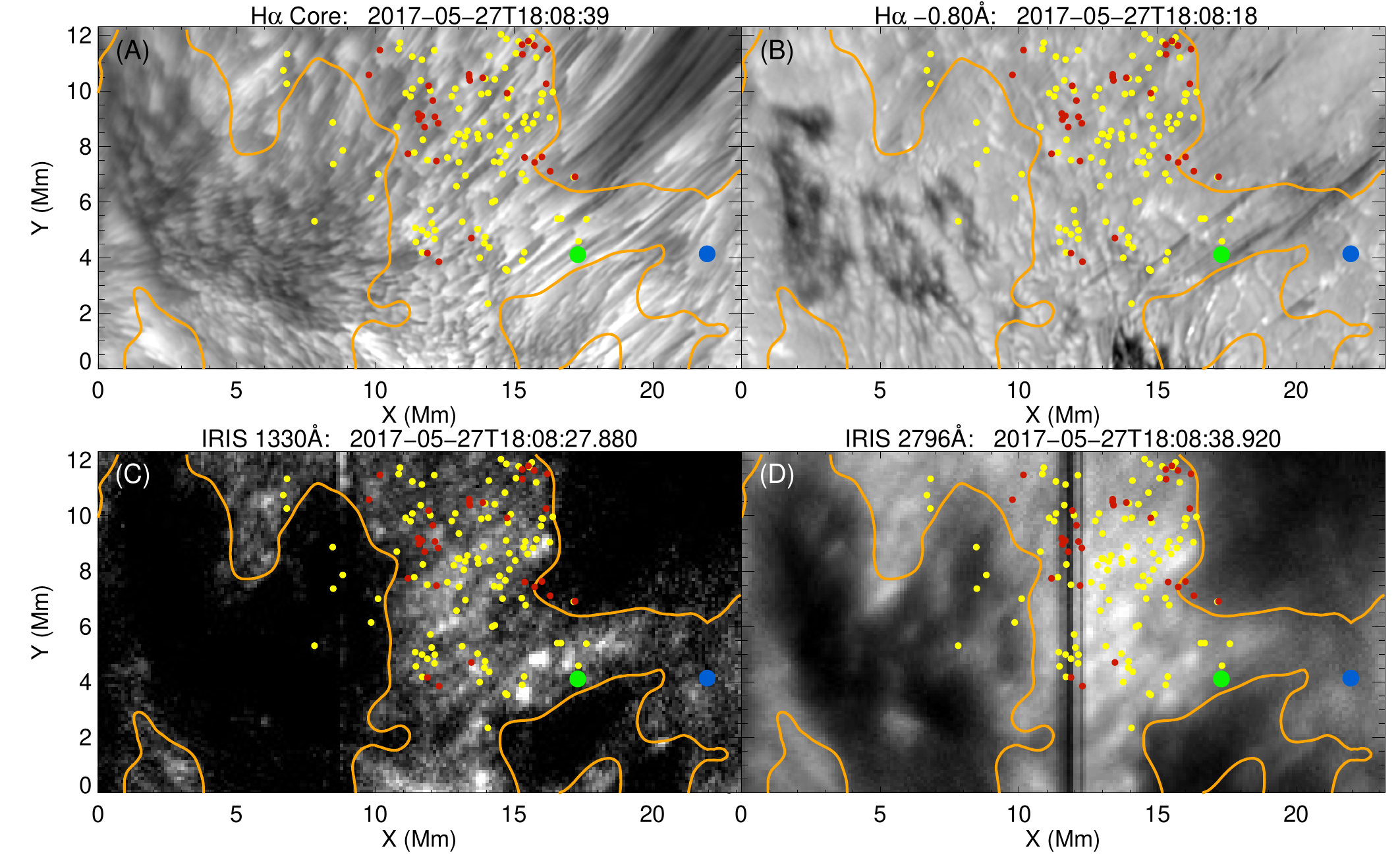}
 \caption{Locations of all the studied SB events in the ROI. (A) and (B): H$\alpha$ core and H$\alpha$ -0.80~{\AA} images respectively. (C) and (D): SJI 1330~{\AA} and 2796~{\AA} images for the same instance. The orange contour, derived from the smoothed SJI 1330~{\AA}, represents the boundary of the moss region in SJI 1330~{\AA}. Small yellow dots show the locations of the SB events seen in SJI 1330~{\AA}, which are related to the spicules in the chromosphere. Small red dots show the locations of the SB events that do not match with any discernible feature in H$\alpha$. The large green dot is the location of the SB event related to the falling plasma. The large blue dot represents the location of the SB event caused by the EB.}
 \label{fig:SBE_loc}
\end{figure}

Higher-resolution observations of the TR have revealed a plethora of brightening events originating from different shapes of magnetic reconnection at different heights, through shocks, etc. \citep{2014_tianb,Skogsrud_2015}. Using coordinated observations of GST and IRIS, we have examined a part of an active region plage/moss to understand the generation mechanisms of these brightening events. We have investigated 131 TR SB events in the region and analyzed their dynamical and multi-wavelength behavior. Figure \ref{fig:SBE_loc} illustrates the locations of these 131 TR SB events. Our analysis reveals that 98 SB events show spatial and temporal correspondence with spicules in H$\alpha$ images; their locations are shown as yellow dots in this figure. Most of these SBs in SJI 1330~{\AA} appear at the tops of the spicules,
 while most of the SBs in 2796~{\AA} images appear to map the body of the spicules. These observations show that the majority of SB events in the plage regions accompany spicules.
 These spicules typically exhibit broadened H$\alpha$ spectral profiles and some of them (9 events) demonstrate parabolic trajectories in the space-time diagrams reminiscent of dynamic fibrils. \cite{2016ApJ...817..124S} observed that a substantial number of SB events in the plage regions are driven by shocks in the chromosphere, producing dynamic fibrils that appear as jets near the H$\alpha$ line core.
In contrast, our statistical investigation reveals that most spicules associated with SB events do not show parabolic trajectories in the space-time diagrams.
\cite{2022MNRAS.515.2672V} indicated a spatiotemporal correspondence between RBEs and SB events in the active regions. However, our detailed analysis shows that the spicules we observe typically do not follow RBE-like H$\alpha$ spectral profiles.

We also observe an event where an EB in H$\alpha$ images appears very close to the TR SBs and shows a temporal match with them in SJI 1330~{\AA} and 2796~{\AA} images. The location of this event is shown as a large blue dot in Figure \ref{fig:SBE_loc}. \cite{2019ApJ...875L..30C} suggested that if the magnetic reconnection happens well above the temperature minimum region, an EB and a nearby UV burst can have a similar underlying generation mechanism.

Furthermore, our analysis also reveals a correspondence between the stream of falling plasma and subsequent brightenings near its footpoint with EB-like H$\alpha$ spectral profiles. The approximate location of these falling-plasma-related-brightening events is shown as a large green dot in Figure \ref{fig:SBE_loc}. The appearance of EB-like H$\alpha$ profiles near the SBs and their temporal match suggest that these have a common origin. Therefore, we report that EB-like H$\alpha$ profiles can be created due to a falling stream of plasma.
Our analysis shows that the speed of the downflows is around 15-20 km s$^{-1}$. Please note that these speeds are measured in the plane of the sky, and the actual speed can be much higher. Furthermore, these downflows are clearly seen in the H$\alpha$ +1 ~{\AA} line position, which is equivalent to around 45 km s$^{-1}$ of Doppler shift. The falling plasma with these high speeds is enough to create termination shocks in the chromosphere, which might generate transient brightenings in TR SJIs and H$\alpha$ far wings due to shock heating. The strong H$\alpha$ wing brightenings associated with falling plasma are also reported by \cite{2016NatSR...624319J}, where a stream of falling plasma along the flare loops creates intense brightenings in the H$\alpha$ at the ending of the stream. They suggested that when siphon-like plasma flows reach supersonic velocities, they could form a shock wave in the atmosphere and cause brightenings.
We need further investigations to uncover their true generation mechanism.

The red dots in Figure \ref{fig:SBE_loc} represent 31 SB events with no discernible chromospheric counterparts. Some of these events are localized intensity enhancements, which might correspond to explosive events. This can be confirmed using spectroscopic analysis, which is out of the scope of this study. However, only a slight difference is observed in the physical properties (see Fig.~\ref{fig:histo}), which may indicate that they might also correspond to even thinner or less energetic spicules that are not detected in the H$\alpha$ chromospheric images due to observational limits.  It is also possible that  these SBs may be connected to other higher atmospheric phenomena (e.g., coronal rain) that are not studied in this article.
It is worth mentioning that some spicules in the chromosphere do not show heating, and as a result, they do not have spatial and temporal correspondence with any SB events in TR SJIs. Some of these spicules could form in regions where the plasma density is relatively high, and the energy is not sufficient to cause heating up to TR temperatures.\\

\section{Conclusions}

Our analysis showed that TR SBs observed in IRIS passbands can be attributed to various chromospheric phenomena, including spicules, EBs, and falling plasma streams. Specifically, we found that  a significant portion of regular SBs (98 of 131) accompany spicules.
These spicules exhibit properties similar to dynamic fibrils and are likely induced by chromospheric shocks, which impact the transition region. Furthermore, we identify SB events associated with EBs, which exhibit significantly higher intensity and longer lifetimes compared to typical SBs in the studied moss region. EBs are believed to stem from magnetic reconnection, and the extended current sheet from this reconnection can trigger brightenings in the TR images.
Our investigation also reveals that in certain areas, SBs may arise from falling plasma streams, producing compact and intense events likely due to termination shocks. The H$\alpha$ line profiles in these regions resemble those of EBs, indicating that falling plasma streams can also generate mustache-like H$\alpha$ profiles.
Additionally, we also found several brightenings (31 of 131) that do not have a clear chromospheric counterpart. We speculate that these events might be generated independently in the TR or may be linked to other higher atmospheric phenomena that are beyond the scope of this study. It is also possible that these brightenings are smaller versions of spicule-related events, with spicules themselves potentially going unnoticed in H$\alpha$ images due to current observational constraints.  These findings suggest a diverse range of mechanisms contributing to brightening events occurring in the transition region, highlighting the complexity of solar atmospheric dynamics.
\\

B.R. thanks the Visting Students Internship Programme (VSP) of the Indian Institute of Astrophysics (IIA) for supporting his internship at IIA. We gratefully acknowledge the use of data from the Goode Solar Telescope (GST) of the Big Bear Solar Observatory (BBSO). BBSO operation is supported by US NSF AGS-2309939 and AGS-1821294 grants and the New Jersey Institute of Technology. GST operation is partly supported by the Korea Astronomy and Space Science Institute and the Seoul National University. IRIS is a NASA small explorer mission developed and operated by LMSAL with mission operations executed at NASA Ames Research Center and major contributions to downlink communications funded by ESA and the Norwegian Space Centre. B.R. and L.P.C. gratefully acknowledge funding by the European Union (ERC, ORIGIN, 101039844). Views and opinions expressed are however those of the author(s) only and do not necessarily reflect those of the European Union or the European Research Council. Neither the European Union nor the granting authority can be held responsible for them. A.C.S. received funding from the Heliophysics Division of NASA’s Science Mission Directorate through the Heliophysics Supporting Research (HSR) Program. V.Y. acknowledges support from NASA 80NSSC20K0025, 80NSSC20K1282, and 80NSSC21K1671 grants, and NSF AGS 2309939, 2300341, and AST 2108235, 2114201 grants.

%\bibliographystyle{aasjournal}
%\bibliography{ref}

\end{document}